# A quantitative measure for the organization of a system, Part 1: A simple case

Georgi Y. Georgiev

Department of Natural Science, Assumption College, 500 Salisbury st, Worcester MA, 02476, USA).
e-mail: georgi@alumni.tufts.edu

Summary

Establishing a universal method to measure the organization of a universal system, will allow us to understand the mechanisms of functioning and organizing in general and enable us to design systems which will posses highest level of perfection. Attempts to measure organization have been done using information and entropy. The least action principle which is the basis of all branches of physics has been applied to a biotechnical system, ecosystem, chemical system and reliability theory to explain their functioning. Here we show that the amount of organization is inversely proportional to the physical amount of action in a system. We show the derivation of the expression for organization and apply it in a simple case. The significance of this result is that it will empower all of the natural and social sciences to quantify the organization in the systems that they are studying when it is applied to them.

Introduction

Measuring the level of organization of a system is extremely important because it will provide a long sought and much needed in all sciences criterion to evaluate the level and to study the mechanisms of organization in all natural and artificial systems[1]. Being able to measure the amount of organization in a system will help us in analyzing and designing complex systems in our society for further improvement of our lives.

The study of organized systems became extremely important and active in recent years[2-23], with the increase of the complexity of our society and the advance of all natural and social sciences. The interest in studying complex systems has also increased, promising to be much more important in the future.

Quantifying organization has been attempted using concepts from theory of complexity like information or entropy[2,3,9]. Several papers have been published applying the principle of least action to a chemical system[11], ecosystem[13], and biotechnical system[12].

In this paper our purpose is to develop a model for quantitatively calculating the amount of organization in a general system. The main idea is that the measure of the amount of organization of a system[1] is inversely proportional to the sum of actions of all its elements. Because the actual paths of the elements in a system are determined by the principle of least action[24] this is the most natural measure of the state of a system. The paths of the elements take into account the configuration of the constraints in the system, so we do not need to explicitly include the state of the constraints in the calculations.

Here we derive an expression for the organization of a system, and then apply it for a simple limiting case of one element and one constraint as an illustration. In Parts 2 and 3 of this paper we will discuss extensions of the model for two and arbitrary number of elements respectively, and will follow the time dependency of the organization in closed and open systems. We have stated the limitations and applicability of this model in the list of assumptions.

## Model

In a previous paper[1] we have stated that for an organized system we can find the natural state of that system as being the one in which the variation of the sum of actions of all of the elements is zero:

$$\delta \sum_{i=1}^{n} I_i = \delta \sum_{i=1}^{n} \int_{t_1}^{t_2} L_i dt = 0 \qquad (1)$$

Where $\delta$ is infinitesimally small variation in the action integral $I_i$ of the $i^{th}$ element, $L_i$ is the Lagrangian of the $i^{th}$ element, and $n$ represents the number of elements in the system, $t_1$ and $t_2$ are the initial and final times of the motions.

In this paper we apply this model is the simplest possible example of a closed system in 2 dimensions with only one element and one internal constraint. The constraint is anything that can change the action integral of an element. We define the boundaries of the closed system (they can be considered as external constraints) to form a square. The positions of the constraints do not enter in the calculations because they are reflected in the length of the trajectory of the element.

List of assumptions:

1. The elements are free particles, not subject to any forces, so the potential energy is a constant and can be set to be zero because the origin for the potential energy can be chosen arbitrary, therefore $V = 0$.

Then, the Lagrangian L of the element is equal only to the kinetic energy $T = \dfrac{mv^2}{2}$ of that element:

$$L = T - V = T = \frac{mv^2}{2} \qquad (2)$$

Where $m$ is the mass of the element, and $v$ is its speed.

2. We are assuming that there is no energy dissipation in this system, so the Lagrangian of the element is a constant:

$$L = T = \frac{mv^2}{2} = \text{constant} \qquad (3)$$

3. The mass *m* and the speed *v* of the element are assumed to be constants.

4. The start point and the end point of the trajectory of the element are fixed at opposite corners of a square [Fig. 1]. This produces the consequence that the action integral cannot become zero, because the end points cannot get infinitely close together.

$$I = \int_{t_1}^{t_2} L\,dt = \int_{t_1}^{t_2} (T-V)\,dt = \int_{t_1}^{t_2} T\,dt \neq 0 \tag{4}$$

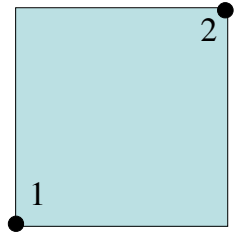

Fig. 1. On this figure we present the boundaries of the closed system, and define the initial 1 and final 2 point of the motion of an element.

5. The constraint cannot block completely the path of the element.

This leads to another consequence that the action integral cannot become infinity, i.e. the trajectory cannot become infinitely long.

$$I = \int_{t_1}^{t_2} L\,dt = \int_{t_1}^{t_2} (T-V)\,dt = \int_{t_1}^{t_2} T\,dt \neq \infty \tag{5}$$

6. The element does not interact with the constraint.

7. In each configuration of the system, the actual trajectory of the element is determined as the one with least action from the Hamilton's Principle[24].

$$\delta I = \int_{t_1}^{t_2} L\,dt = \int_{t_1}^{t_2} (T-V)\,dt = \int_{t_1}^{t_2} T\,dt = 0 \tag{6}$$

8. The medium inside the system is isotropic (it has all its properties identical in all directions). The consequence of this assumption is that the constant velocity of the element allows us to substitute the interval of time with the length of the trajectory of the element, as we will see below.

Building the model:

1. The organization is proportional to the inverse of the sum of actions of all elements.

$$Org \propto \frac{1}{\sum_{i=1}^{n} I_i} \qquad (7)$$

   For this model we set n=1, then:

$$Org \propto \frac{1}{I} \qquad (8)$$

2. For the purpose of this model, we accept the numerator to be a constant, so up to this constant we set the organization for a closed system to be equal to the reciprocal of the sum of the actions of all of the fixed number of elements in the system. For simplicity we set this constant to be equal to one.

$$Org = \frac{Const}{\sum_{i=1}^{n} I_i} = \frac{1}{\sum_{i=1}^{n} I_i} \qquad (9)$$

Where the quantity of organization of the system is abbreviated to Org. For n=1 we have:

$$Org = \frac{Const}{I} = \frac{1}{I} \qquad (10)$$

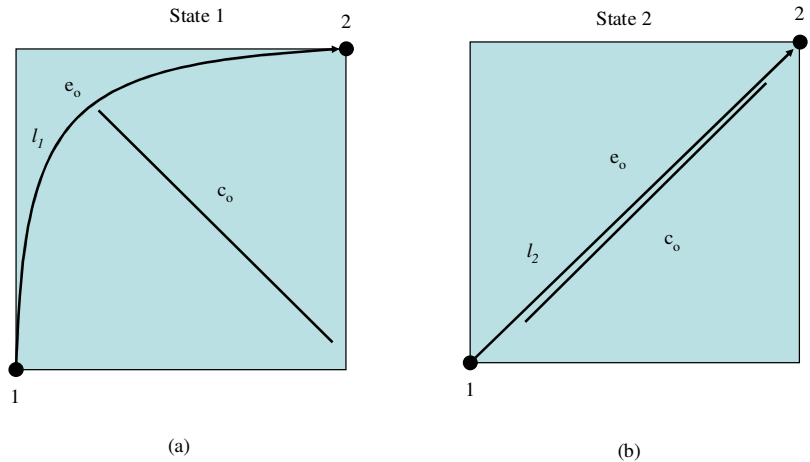

Fig. 2 This is a cartoon representation of the two states of the system with a constraint noted here by $C_o$, and the least action path of the element noted here by $e_o$ in each case as it can be found from Hamilton's Principle[2] eq. 6. Here $l_1$ and $l_2$ are the length of the trajectory of the element in each case.

The question that we want to answer in this model is which of the two states on Fig. 2 is better organized. In the first state (Fig. 2 (a) ) the element has its trajectory of least action in the given configuration of the constraint to be twice the length of its trajectory in the second case (Fig. 2 (b) ), because of the specific orientation of the constraint. We calculate the organization of the system in each state by measuring the action in both cases and applying eq. 10 to each of them.

Under the assumptions of this model, the kinetic energy is a constant (assumption #2), so the action integral accepts the form:

$$I = \int_{t_1}^{t_2} L dt = \int_{t_1}^{t_2} T dt = T(t_2 - t_1) = T \Delta t = \frac{mv^2}{2} \Delta t = C_1 \Delta t \qquad (11)$$

Where $C_1 = T = \frac{mv^2}{2}$ and $\Delta t$ is the interval of time that the motion of the element takes.

An approximation for isotropic medium (assumption 8) allows us to express the time through the speed of the element when it is constant (Assumption 3). In this case then we can solve $v = \frac{l}{\Delta t}$ which is the definition of average velocity for the interval of time as $\Delta t = \frac{l}{v}$, where $l$ is the length of the trajectory of the element in each case from p.1 to p.2.

The speed of the element $v$ is fixed to be another constant, so the action integral accepts the form:

$$I = C_1 \Delta t = C_1 \frac{l}{v} = C_2 l \qquad (12)$$

Where $C_2 = \frac{C_1}{v}$.

When we substitute eq. 12 in the expression for organization, eq. 10, we obtain:

$$Org = \frac{Const}{I} = \frac{1}{I} = \frac{1}{C_2 l} = \frac{C_3}{l} \qquad (13)$$

Where $C_3 = \frac{1}{C_2}$ is another constant.

Now we turn to the two states of the system with different orientation of the constraint and different action of the elements, as shown on Fig. 2. The organization of those two states is respectively: $Org_1 = \frac{C_3}{l_1}$ in state 1, and $Org_2 = \frac{C_3}{l_2}$ in state 2 of the system. On fig. 2, the length of the trajectory in the second case (b) is less, $l_2 < l_1$ which produces the result that state 2 is the state with a better organization. The difference between the organizations in the two states of the same system is in general:

$$Org_2 - Org_1 = \frac{C_3}{l_2} - \frac{C_3}{l_1} = C_3 \left( \frac{1}{l_2} - \frac{1}{l_1} \right) = C_3 \left( \frac{l_1 - l_2}{l_1 l_2} \right) \qquad (14)$$

This can be rewritten as:

$$\Delta Org = C_3 \left( \frac{\Delta l}{\prod_{i=1}^{2} l_i} \right) \qquad (15)$$

Where $\Delta Org = Org_2 - Org_1$, $\Delta l = l_1 - l_2$ and $\prod_{i=1}^{2} l_i = l_1 l_2$.

### An example

If $l_1 = 2 l_2$, or the first trajectory is twice as long as the second this expression produces the result: $Org_1 = \frac{C_3}{2l_2} = \frac{Org_2}{2}$ or $Org_2 = 2 \times Org_1$ or that state 2 is twice better organized than state 1.

Alternatively, substituting in eq. 15 we have, $Org_2 - Org_1 = C_3 \left( \frac{2-1}{2} \right) = \frac{C_3}{2}$ or there is 50% difference between the two organizations, which is the same as to say that the second of them is quantitatively twice as well organized than the first one. This example illustrates the purpose of the model for direct comparison between the amounts of organization in two different states of a system given that all conditions other than the orientation of the constraint are unchanged.

### Discussion

The eq. 15 allows calculation of the organization in a system. The method uses the most fundamental principle in Physics, the principle of least action which describes the actual motions of objects. In an example of one element and one constraint, a state of the system where the element has half of the action than in another state the system is calculated to have double of the amount of organization. The model in this paper is limited only to closed, nondissipative system with one free particle as an element, and one constraint. An extension of the model to open systems of $n$ elements will provide a method for calculation of the level of organization of any system. The significance of this result is that it provides a quantitative measure for comparing different levels of organization in the same system, and for comparing different systems. Measuring the organization of systems should facilitate the development of all natural and social sciences.